\documentclass[aps,preprint]{revtex4}%
\usepackage{amsfonts}
\usepackage{amsmath}
\usepackage{amssymb}
 \usepackage{subcaption}
\usepackage{graphicx}
 \usepackage{subfig,caption}
\usepackage{mathtools}
\providecommand{\U}[1]{\protect\rule{.1in}{.1in}}

\begin{document}
\title{Charged particle geodesics and closed timelike curves in an electromagnetic universe}
\author{M. Halilsoy}
\email{mustafa.halilsoy@emu.edu.tr}
\author{V. Memari}
\email{vahideh.memari@emu.edu.tr}
\affiliation{Department of Physics, Faculty of Arts and Sciences, Eastern Mediterranean
University, Famagusta, North Cyprus via Mersin 10, Turkey}
\date{\today}

\begin{abstract}
The spinning electromagnetic universe, known also as the Rotating Bertotti-Robinson(RBR) spacetime is considered as a model to represent our cosmos. The model derives from different physical considerations, such as colliding waves, throat region, and near horizon geometry of the Kerr-Newman black hole. Our interest is whether such a singularity-free spinning cosmology gives rise to a natural direction of flow, a 'chirality' for charged particles. Homochiral structures are known to be crucial for biology to start. Our concern here is cosmology rather than biology, but as in biology, the stable structures in cosmology may also rely on homochiral elements. We show the occurrence of closed timelike curves a 'la' G{\"o}del. Such curves, however, seem possible only at localized cell structures, not at large scales, but according to our prescription of near horizon geometry, they arise in the vicinity of any charged, spinning black hole.
\end{abstract}
\maketitle

\section{Introduction}
The anisotropic extension of the Bertotti-Robinson (BR) \cite{1,2} electromagnetic (em) universe, known as the Rotating Bertotti-Robinson (RBR)\cite{3} is considered a model with interesting physical properties. It is free of singularities and admits conformal curvature due to rotation. We investigate charged particle geodesics and the possibility of closed timelike curves in RBR. This spacetime was known first as the 'throat' region between rotating black holes (BHs)\cite{3}. It was shown later on that the same spacetime emerges as a result of two colliding em waves with cross polarization\cite{4,5}. Further, it can be shown that the RBR spacetime arises (in tuned parameters) as a near horizon geometry (NHG) of the extremal Kerr-Newman (KN) BH.\cite{6,7}. With this much characterization, it is natural to expect that a proportion of our universe is governed by such a model. Its em field consists of polarized electric ($\Vec{E}$) and magnetic fields ($\Vec{B}$) originating from the rotation of the entire universe. In particular, the existence of polarized $\Vec{B}$ fields deserves more attention, which make the principal aim of this article.\par
To motivate our readers we recall a well-known problem from classical mechanics. On our rotating Earth a northward moving car deflects to the right in the northern hemisphere and to the left in the southern hemisphere. This is due to the active Coriolis face \cite{8} from the rotation of Earth. Similar behavior occurs for oppositely charged particles in a fixed magnetic field. Opposite charges deflect differently also in the $\Vec{B}$ field of our RBR cosmology. Can this be interpreted as a natural 'chirality' for charged objects in the cosmos? Interestingly molecular chirality was observed first by L. Pasteur in 1848. In other words, cosmological $\Vec{B}$ may create a natural chirality for charged systems. For stable systems, homochiral molecules/systems must stick together in analogy with the spinning particles. And oppositely chiral systems form unstable systems ready to split. Although to date a complete answer is not known to the problem of chirality in the cosmos the magnetic field in the model universe that we undertake to investigate, namely the RBR, may contribute at a cosmological level. The $\Vec{E}$ and $\Vec{B}$ fields are polarized by the existence of the gravitational field. The em and gravitational invariants are both dependent only on the angular variable $\theta$, which suggests that the $\Vec{B}$ field gives rise at every point, irrespective of the coordinate $r$, $a$ whirlpool structure. Such formations are absent when the $\Vec{B}$ field vanishes. The phase factor of the em field spinor field $\Phi_{1}\left(\theta\right)$ is the determining factor for the $\Vec{B}$ field and also the cross-polarization mode. The cosmic microwave background (CMB) web patterns also may be connected with the second mode of the $\Vec{B}$ field. In our previous study of geodesic analysis, we took into account only the linear mode in the absence of the $\Vec{B}$ field \cite{5}. In the present paper, we intend to study geodesics thoroughly with the $\Vec{B}$ field. \par
Our second concern in this article is the formation of closed timelike curves (CTCs). Since the seminal work f K.G{\"o}del \cite{9} where he had shown that a rotating cosmological fluid spacetime admits CTCs, there have been numerous attempts to do the same in various spacetimes. It is known by now that each spinning spacetime is a potential candidate for this purpose. The exact solutions of Einstein's equations that admit a cross-term in the metric such as Kerr, NUT, and their derivatives are all in this category. More recently we applied the same argument to a more tedious stationary, charged metric \cite{10}. The main idea in this approach is to project the 4D manifold to the $\left(t,\varphi\right)$ sector, and take the time proportional to $\varphi$ which is periodic so that the time itself becomes periodic. This procedure creates CTCs so that after each period an object returns to the starting point. We apply the idea of G{\"o}del to the metric that is the subject matter in this article, namely the RBR. In the coordinate system $\left(t,r,\theta,\varphi\right)$ we work out the requirement for the existence of CTCs. We find possible solutions in particular for $\theta=\frac{\pi}{2}$ and $\theta=\frac{\pi}{4}$, which can be extended to any possible angle. Our finding, however, suggests that going backward in time in the RBR cosmology is severely confined to very small scales of distances, which does not happen at large scales. In other words, we have found only small, confined whirlpool cells that admit going backward in time.\par
The organization of the paper is as follows. In section II we revisit the RBR spacetime and summarize its properties. In section III we study geodesics. CTCs follow in section IV, and we conclude the paper with our Conclusion and Discussion in section V.
\section{The RBR spacetime}
RBR spacetime was discovered first by Carter \cite{3}. Later on, it was rediscovered as the spacetime resulting from the collision of two cross-polarized electromagnetic shock waves \cite{5}. In the coordinates $\left(t,r,\theta,\varphi\right)$ the RBR spacetime is given by
\begin{equation}
    ds^{2}=\frac{F\left(\theta\right)}{r^{2}}\left[dt^{2}-dr^{2}-r^{2}d\theta^{2}-\frac{r^{2}\sin^{2}{\theta}}{F^{2}\left(\theta\right)}\left(d\varphi-\frac{q}{r}dt\right)^{2}\right]\label{1}
\end{equation}
where
 \begin{align}
     F\left(\theta\right) & =1+a^{2}\left(1+\cos^{2}\theta\right) \nonumber\\
 q & =2a\sqrt{1+a^{2}}\label{2}
 \end{align}
 in which $0<a<\infty$ is the rotation parameter. It is easily seen that for $a=0$ we recover the BR spacetime, which represents the conformally flat (CF), solution of the EM equations. In addition to the 'throat' property connecting two black holes, we shall show that the RBR arises as a near horizon geometry (NHG) of the Kerr-Newman (KN) spacetime in the extreme limit. To this end, we recall the NHG of the Extremal KN in the form \cite{6,7}
 \begin{equation}
     ds^{2}=\left(1-\frac{a^{2}\sin^{2}{\theta}}{r_{0}^{2}}\right)\left[\left(\frac{d\bar{t}}{r_{0}r}\right)^{2}-\left(\frac{r_{0}dr}{r}\right)^{2}-\left(r_{0}d\theta\right)^{2}\right]-r_{0}^{2}\sin^{2}\theta\left(1-\frac{a^{2}\sin^{2}{\theta}}{r_{0}^{2}}\right)^{-1}\left(d\bar{\varphi}+\frac{2Ma}{rr_{0}^{4}}d\Bar{t}\right)^{2} \label{3}
 \end{equation}
in which $r_{0}=const$, $a=$ rotation parameter of Kerr metric and $\left(\bar{t},\bar{\varphi}\right)$ are shown with over bars since we shall rescale them. By the further identification
\begin{align}
    r_{0}^{2}&=1+2a^{2}\nonumber\\
    \bar{\varphi}&=\frac{\varphi}{r_{0}^{2}}\label{4}\\
    \bar{t}&=-r_{0}^{2}t\nonumber\\
    M&=\sqrt{a^{2}+1}\nonumber
\end{align}
we obtain the RBR (\ref{1}). That is, the line element (\ref{1}) describes also the NHG of the KN geometry in a particular tuning of the parameters.
\subsection{The Invariants of the spacetime}
By choosing the Newman-Penrose (NP) null tetrad \cite{11} basis $1-$ forms
\begin{align}
    l& =\frac{\sqrt{F}}{2\sqrt{2}r}\left(dt-dr\right) \nonumber \\
    n& =\frac{\sqrt{2F}}{r}\left(dt+dr\right) \nonumber \\
    \sqrt{2}m& =i\sqrt{F}d\theta+\frac{\sin \theta}{\sqrt{F}}\left(\frac{q}{r}dt-d\varphi\right)\label{5}\\
    \bar{m}&=\left(\textit{c.c}\right) \textit{of } m\nonumber
    \end{align}
    we obtain the only nonzero NP scalars given by
    \begin{align}
    \Phi_{11}& =\frac{1}{4F^{2}}\nonumber \\
   \Psi_{2}& =\frac{a^{2}}{F^{3}}\left[\left(1+a^{2}\right)\cos{2\theta}+a^{2}\cos^{2}\theta-\frac{i}{a}\sqrt{1+a^{2}}\cos\theta\left(1+2a^{2}+a^{2}\sin^{2}\theta\right)\right]\label{6}
\end{align}
It is seen that these quantities are regular and the invariants $\Phi_{11}$ and $\Psi_{2}$ obtained in the proper null tetrad are independent of $r$ depending only on the angle $\theta$. The rotation parameter creates a non-isotropy in the spacetime. The maximum non-isotropy can be formed by the ratio of $\Phi_{11}\left(\theta=\frac{\pi}{2}\right)$ over $\Phi_{11}\left(\theta=0\right)$, for $a\rightarrow \infty$ which gives $25$ percent.
\subsection{The electromagnetic properties of the spacetime}
The general experiment for the electromagnetic (em) $2-$ form in terms of the Maxwell spinors $\left(\Phi_{0},\Phi_{1},\Phi_{2}\right)$ is given by
\begin{equation}
    F=-\left(\Phi_{1}+\bar\Phi_{1}\right)l\wedge n+\Phi_{2}l\wedge m+\bar\Phi_{2}l\wedge \bar m-\bar\Phi_{0}n\wedge m-\Phi_{0}n\wedge \bar m+\left(\Phi_{1}-\bar\Phi_{1}\right)m\wedge \bar m \label{7}
\end{equation}
in which a 'bar' implies a complex conjugate. The em spinors are given by 
\begin{align}
    \Phi_{0}&=F_{\mu\nu}l^{\mu}m^{\nu}\nonumber\\
    \Phi_{1}&=\frac{1}{2}F_{\mu\nu}\left(l^{\mu}n^{\nu}+\bar{m}^{\mu}m^{\nu}\right)\nonumber\\
    \Phi_{2}&=F_{\mu\nu}\bar{m}^{\mu}n^{\nu}\label{8}
\end{align}
Since we have only $\Phi_{11}\neq 0$, then $F$ reduces to 
\begin{equation}
    F=-\left(\Phi_{1}+\bar\Phi_{1}\right)l\wedge n+\left(\Phi_{1}-\bar\Phi_{1}\right)m\wedge \bar m \label{9}
\end{equation}
and its dual $2-$ form 
\begin{equation}
    \prescript{\ast}{}{F_{\mu\nu}}=-i\left(\Phi_{1}+\bar{\Phi}_{1}\right)m\wedge\bar{m}+i\left(\Phi_{1}-\bar{\Phi}_{1}\right)l\wedge{n}\label{10}
\end{equation}
From the chosen proper null tetrad we have
    \begin{align}
        l\wedge n & =\frac{F}{r^{2}}dt\wedge dr \nonumber \\
        m\wedge\bar m & =i\sin \theta d\theta \wedge \left( \frac{q}{r}dt-d\varphi \right) \label{11}
        \end{align}
and $\Phi_{1}=\frac{1}{2F}e^{i\alpha}$ with the phase function $\alpha$.
Substituting $\Phi_{1}$ into Eq. (\ref{9}) and using (\ref{11}) we obtain the components of the em field tensor $F_{\mu\nu}=\partial_{\mu}A_{\nu}-\partial_{\nu}A_{\mu}$ as follows
\begin{align}
   F_{tr}& =-\frac{\cos\alpha}{r^{2}}\nonumber \\
   F_{t\theta}& =\frac{q\sin\alpha \sin\theta}{rF}\nonumber \\
   F_{\theta \phi}& =\frac{\sin\alpha \sin\theta}{F} \label{12}
\end{align}
These components solve the Maxwell equations with the phase function
\begin{equation}
    \alpha=2\arctan\left(\frac{a}{\sqrt{1+a^{2}}}\cos{\theta}\right)\label{13}
\end{equation}
so that we have
\begin{align}
    \sin{\alpha}&=\frac{q\cos{\theta}}{F}\nonumber\\
    \cos{\alpha}&=\frac{1}{F}\left(1+a^{2}\sin^{2}\theta\right)\label{14}
\end{align}
\begin{figure}[h]
\centering
 \includegraphics[scale=0.4]{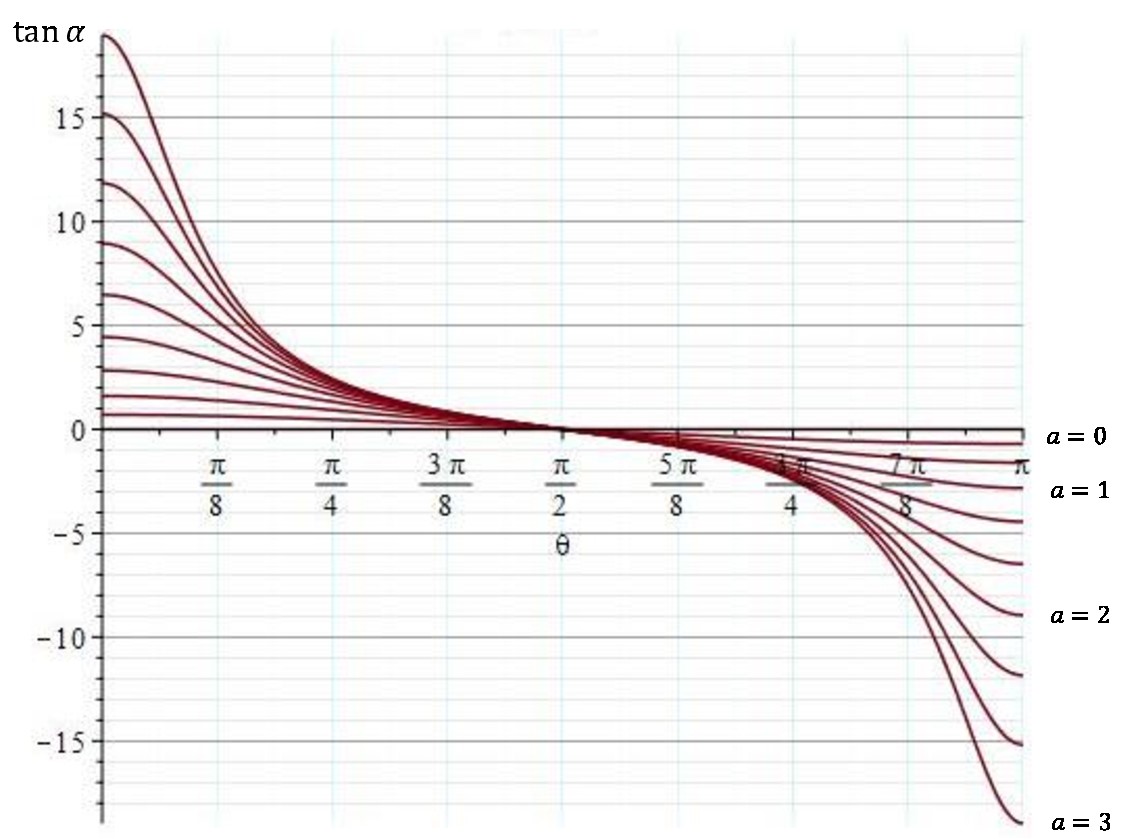}
        \caption{$tan\alpha$ versus $\theta$. The $\tan\alpha$ plot for various spin parameters $a$, from Eq. (\ref{13}), which is the polarization of the $\Vec{B}$ field}
        \label{FIG. 1}
\end{figure}
\begin{figure}
     \centering
     \begin{subfigure}[h]{0.4\textwidth}
         \centering
         \includegraphics[width=\textwidth]{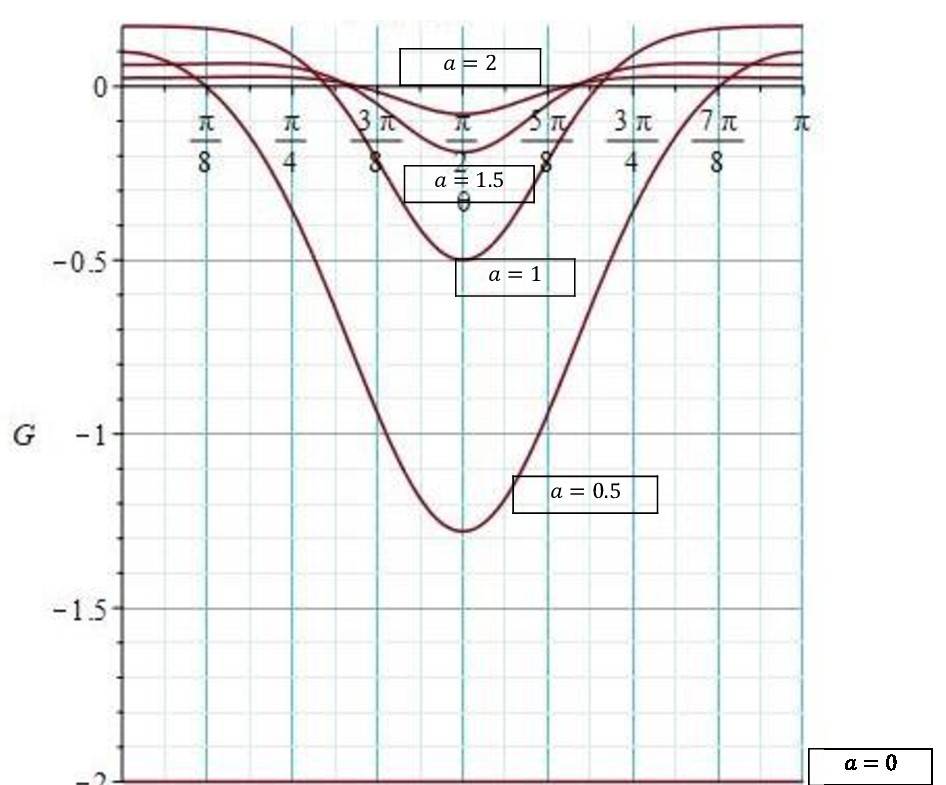}
         \caption{$G$ versus $\theta$ for $0<a<2$}
         \label{Fig. 2a}
     \end{subfigure}
     \hfill
     \begin{subfigure}[h]{0.4\textwidth}
         \centering
         \includegraphics[width=\textwidth]{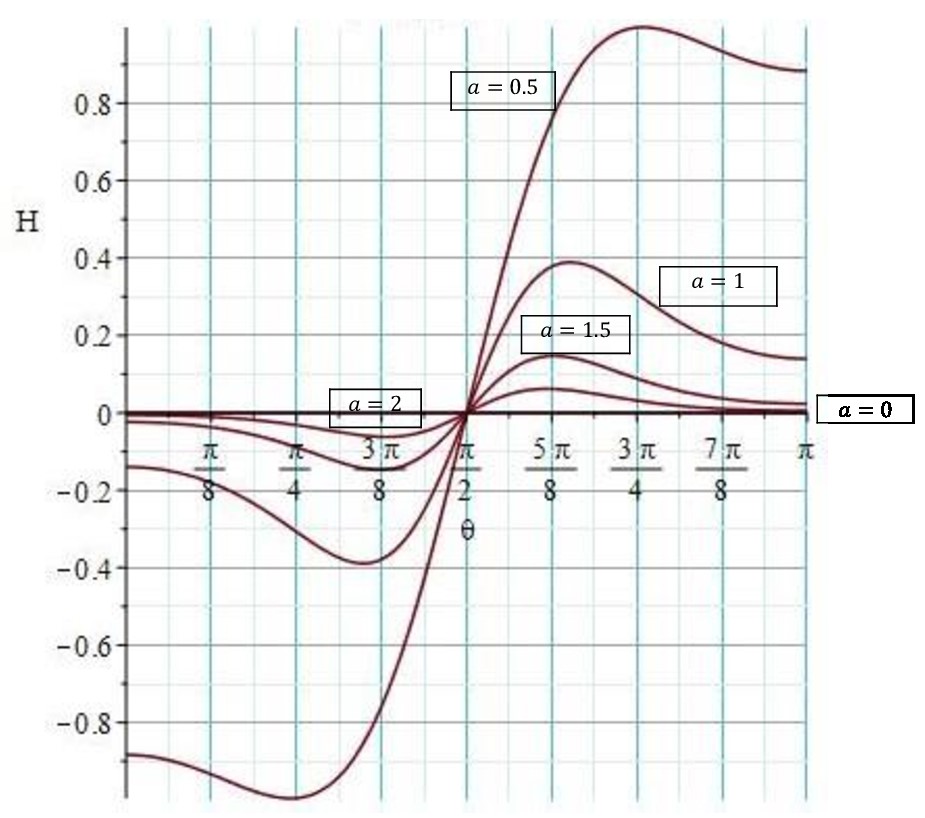}
         \caption{$H$ versus $\theta$ for $0<a<2$}
         \label{Fig. 2b}
     \end{subfigure}
        \caption{Fields invariants plots versus $\theta$ for various constants $a$.}
        \label{FIG. 2}
\end{figure}
In the extreme spinning case ($a\rightarrow \infty$) we have $e^{i\alpha}=\frac{1+i\cos{\theta}}{1-i\cos{\theta}}$. The plot of $\tan\alpha$ versus $\theta$ which is a measure of second polarization is depicted in Fig. (\ref{FIG. 1}). The invariants of em field are easily found from (\ref{9}) and (\ref{10}) as 
\begin{align}
    G\coloneqq & F_{\mu\nu}F^{\mu\nu}  = -\frac{2\cos \left(2\alpha \right)}{F^{2}}\nonumber \\
   H\coloneqq & F_{\mu\nu}\prescript{\ast}{}{F^{\mu\nu}}=-\frac{\sin\left(2\alpha\right)}{F^{2}} \label{15}
\end{align}

which are plotted in Fig.(\ref{FIG. 2}).
\begin{figure}
     \centering
     \begin{subfigure}[t]{0.4\textwidth}
         \centering
         \includegraphics[width=\textwidth]{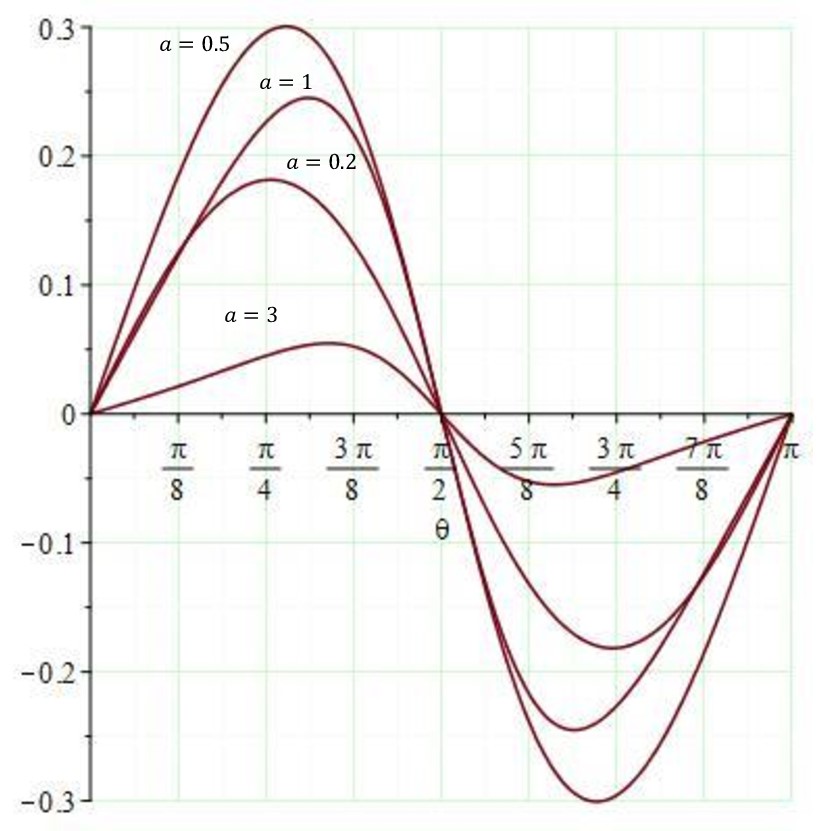}
         \caption{$F_{\theta\varphi}$ versus $\theta$ for $a=1,3,0.2,0.5$}
         \label{Fig. 3a}
     \end{subfigure}
     \hfill
     \begin{subfigure}[t]{0.4\textwidth}
         \centering
         \includegraphics[width=\textwidth]{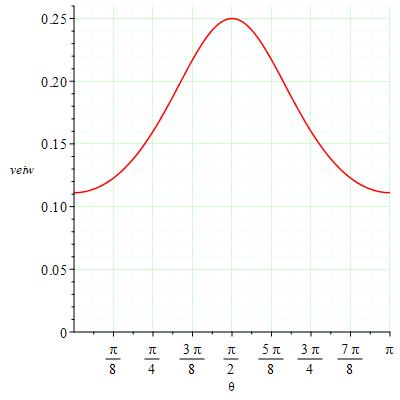}
         \caption{$\Phi_{11}$ versus $\theta$ for $a=1$}
         \label{Fig. 3b}
     \end{subfigure}
        \caption{The magnetic field and energy-momentum (Ricci) plots versus $\theta$}
        \label{FIG. 3}
\end{figure}
It can easily be observed that the choice $\alpha=0$, removes the magnetic field which leaves us only with the electric component $F_{tr}=-\frac{1}{r^{2}}$. Geodesic analysis that was worked out long ago considered only this particular case \cite{5}. Our intention now is to consider the general analysis with a non-zero magnetic field. $F_{\theta\varphi}$ component represents the magnetic field, if we rewrite $F_{\theta\varphi}$ of Eq. (\ref{12}) in terms of $\theta$ we get
\begin{equation}
F_{\theta\varphi}=\frac{a\sqrt{1+a^{2}}\sin\left(2\theta\right)}{\left[1+a^{2}\left(1+\cos^{2}\theta\right)\right]^{2}}\label{16} 
\end{equation}
For various $a$ parameters, magnetic field and Ricci component $\Phi_{11}$ are plotted in Fig. (\ref{FIG. 3}). This expression suggests that the strength of the $\Vec{B}$ field is inversely proportional to the spin parameter of the central BH. For $a\rightarrow \infty$, $F_{\theta\varphi}\rightarrow 0$ and the em field is ruled only by the $\Vec{E}$ field. For very weak rotations ($a<<1$), we have the non-zero em components $F_{tr}\approx - \frac{1}{r^{2}}$, $F_{t\theta}\approx 0$ and  $F_{\theta\varphi}\approx a \sin{\theta}\cos{\theta}$.
\section{Geodesics of charged particles}
The Lagrangian of a charged particle with charge $Q$ is described as
\begin{equation}
\mathcal{L}=\frac{1}{2}g_{\mu \nu}\overset{\cdot}{x}^{\mu}\overset{\cdot}{x}^{\nu}+QA_{\mu}\overset{\cdot}{x}^{\mu}  \label{17}
\end{equation}
metric of RBR spacetime is defined in Eq. (\ref{1}) and vector potential $A_{\mu}$ can be found from Eq. (\ref{12}) as follows
\begin{align}
    A_{t}&=-\frac{\cos \alpha}{r}\nonumber \\
    A_{\varphi}&=\frac{q}{2a^2F}\label{18}
\end{align}
By replacing Equations (\ref{1}) and (\ref{18}) in Lagrangian (Eq. \ref{17}) we obtain
\begin{equation}
\mathcal{L}=\frac{F}{2r^{2}}\left[\overset{\cdot}{t}^{2}-\overset{\cdot}{r}^{2}-r^{2}\overset{\cdot}{\theta}^{2}-\frac{r^{2}\sin^{2}\theta}{F^{2}}\left(\overset{\cdot}{\varphi}-\frac{q}{r}\overset{\cdot}{t}\right)^{2}\right]-\frac{Q\left(1+a^{2}\sin^{2}\theta\right)}{rF}\overset{\cdot}{t}+\frac{Qq}{2a^{2}F}\overset{\cdot}{\varphi}  \label{19}
\end{equation}
Geodesics equations for this Lagrangian is found
\begin{align}
    &\frac{F\overset{\cdot}{t}}{r^{2}}+\frac{q\sin^{2}\theta}{rF}\left(\overset{\cdot}{\varphi}-\frac{q}{r}\overset{\cdot}{t}\right)-\frac{Q\left(1+a^{2}\sin^{2}\theta\right)}{rF}=E=const.\nonumber \\
    &-\frac{\sin^{2}\theta}{F}\left(\overset{\cdot}{\varphi}-\frac{q}{r}\overset{\cdot}{t}\right)+\frac{Qq}{2a^{2}F}=-l=const.\nonumber \\
    &\frac{F^{/}}{2r^{2}}\left[\overset{\cdot}{t}^{2}-\overset{\cdot}{r}^{2}-r^{2}\overset{\cdot}{\theta}^{2}-\frac{r^{2}\sin^{2}\theta}{F^{2}}\left(\overset{\cdot}{\varphi}-\frac{q}{r}\overset{\cdot}{t}\right)^{2}\right]-\frac{F}{2}\left(\frac{\sin^{2}\theta}{F^{2}}\right)^{/}\left(\overset{\cdot}{\varphi}-\frac{q}{r}\overset{\cdot}{t}\right)^{2}-\frac{Q}{r}\left(\frac{1+a^{2}\sin^{2}\theta}{F}\right)^{/}\overset{\cdot}{t}+\nonumber\\
    &+\frac{Qq}{2a^{2}}\left(\frac{1}{F}\right)^{/}\overset{\cdot}{\varphi}-\frac{d}{d\tau}\left(-F\overset{\cdot}{\theta}\right)=0\label{20}\\
    &\frac{F}{r^{2}}\left[\overset{\cdot}{t}^{2}-\overset{\cdot}{r}^{2}-r^{2}\overset{\cdot}{\theta}^{2}-\frac{r^{2}\sin^{2}\theta}{F^{2}}\left(\overset{\cdot}{\varphi}-\frac{q}{r}\overset{\cdot}{t}\right)^{2}\right]=1\nonumber
\end{align}
From now on to simplify our calculations we consider the plane $\theta=\frac{\pi}{2}$ so that geodesics equations shall reduce as follows
\begin{align}
    &\overset{\cdot}{t}=\frac{r^{2}}{F}\left[E-\frac{C_{0}q}{rF}+\frac{Q}{r}\right]\label{21} \\
    &\overset{\cdot}{\varphi}-\frac{q}{r}\overset{\cdot}{t}=C_{0}=const.\label{22}\\
    &\overset{\cdot}{r}^{2}=\frac{r^{4}}{F^{2}}\left(E+\frac{1}{r}\left(Q-\frac{C_{0}q}{F}\right)^{2}\right)-\frac{r^{2}}{F^{2}}\left(C_{0}^{2}+F\right)\label{23}
\end{align}
where $C_{0}=F\left(\frac{Qq}{2a^{2}F}+l\right)$ and $F=1+a^{2}$ a constant parameter depending on rotation factor. We can rewrite the Eq. (\ref{21}) and Eq. (\ref{23}) by defining a new constant $B_{0}=Q-\frac{qC_{0}}{F}$:
\begin{align}
    &\overset{\cdot}{t}=\frac{r^{2}}{F}\left[E+\frac{B_{0}}{r}\right]\label{24} \\
    &\overset{\cdot}{r}^{2}=\frac{r^{4}}{F^{2}}\left(E+\frac{B_{0}}{r}\right)^{2}-\frac{r^{2}}{F^{2}}\left(C_{0}^{2}+F\right)\label{25}
\end{align}
Equation (\ref{25}) is solvable and we can obtain the exact solution for $r\left(\tau\right)$ for both cases $k^{2}>0$ and $k^{2}<0$.\\
For $k^{2}>0$:
\begin{equation}
    r\left(\tau\right)=\frac{1}{z_{0}\cos{k\tau}+\frac{B_{0}E}{k^{2}F^{2}}}\label{26}
\end{equation}
For $k^{2}<0$:
\begin{equation}
    r\left(\tau\right)=\frac{1}{z_{0}\cosh{k\tau}+\frac{B_{0}E}{k^{2}F^{2}}}\label{27}
\end{equation}
which $z_{0}$ is an integration constant and $k^{2}=\frac{1}{F^{2}}\left(C_{0}^{2}+F-B_{0}^{2}\right)$. Dividing Eq. (\ref{25}) to Eq. (\ref{24}) gives us
\begin{equation}
    \left(\frac{dr}{dt}\right)^{2}=1-\frac{F+C_{0}^{2}}{\left(Er+B_{0}\right)^{2}}\label{28}
\end{equation}
\begin{figure}
     \centering
     \begin{subfigure}[b]{0.4\textwidth}
         \centering
         \includegraphics[width=\textwidth]{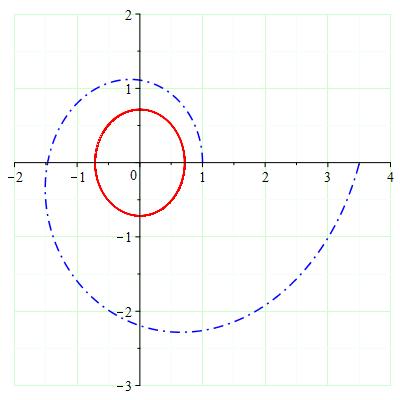}
         \caption{numeric plot of $r$ versus $\varphi$ for positive charge,$Q=2$}
         \label{Fig. 4a}
     \end{subfigure}
     \hfill
     \begin{subfigure}[b]{0.4\textwidth}
         \centering
         \includegraphics[width=\textwidth]{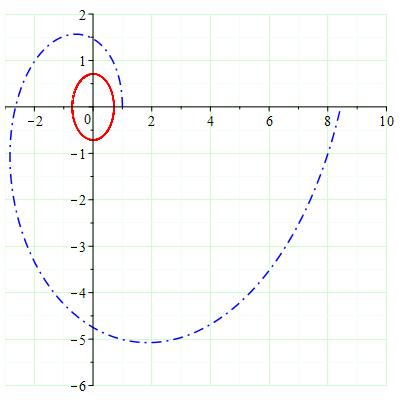}
         \caption{numeric plot of $r$ versus $\varphi$ for negative charge,$Q=-2$}
         \label{Fig. 4b}
     \end{subfigure}
        \caption{Numeric plot of $r$ versus $\varphi$ for both positive and negative charge is indicated. For simplicity, we choose constants, $l=1$, $Q= \pm2$, and rotation parameter $a=1$ with the initial point $r_{0}=1$ from where the particle is released. Starting from rest gives us the initial condition for a positive charge $E=9.859700171$ and for the negative charge $E=2.467418127$. For the specific case of $a=5$ with the same constant values at $r=0.7153196154$ for $Q=2$, and $r=0.7106719373$ for a negative charge, we obtain circular motion for the charged particle (red circle). Also, it is possible to find circular orbits for different values of $a$ or $Q$. For $a=1$ and $a=5$ the right-handed behavior of both $\pm$ charges can be observed}
        \label{FIG. 4}
\end{figure}
then acceleration will be 
\begin{equation}
    \left(\frac{d^{2}r}{dt^{2}}\right)=\frac{2E\left(F+C_{0}^{2}\right)}{\left(Er+B_{0}\right)^{3}}\label{29}
\end{equation}
the sign of acceleration depends on $C_{0}$ and $B_{0}$ constants, in other words, the sign of the particle's charge. After replacing Eq. (\ref{24}) in Eq. (\ref{22})and doing the same process for Eq. (\ref{25}) and Eq. (\ref{22}) we get
\begin{equation}
    \left(\frac{dr}{d\varphi}\right)^{2}=\frac{r^{2}}{\left[C_{0}F+q\left(Er+B_{0}\right)\right]^{2}}\left[\left(Er+B_{0}\right)^{2}-\left(C_{0}^{2}+F\right)\right]\label{30}
\end{equation}
Upon differentiating Eq. (\ref{30}) we obtain
\begin{equation}
   \frac{d^{2}r}{d\varphi^{2}}=\frac{r\left[\left(Er+B_{0}\right)^{2}-\left(C_{0}^{2}+F\right)\right]}{\left[C_{0}F+q\left(Er+B_{0}\right)\right]^{2}}-\frac{qEr^{2}\left[\left(Er+B_{0}\right)^{2}-\left(C_{0}^{2}+F\right)\right]}{\left[C_{0}F+q\left(Er+B_{0}\right)\right]^{3}}+\frac{Er^{2}\left(Er+B_{0}\right)}{\left[C_{0}F+q\left(Er+B_{0}\right)\right]^{2}}\label{31}
\end{equation}
Analytically this is solvable for $r\left(\varphi\right)$, however, we shall be satisfied with the numerical plot of charged particle trajectories with the initial condition  $\left(\frac{dr}{d\varphi}\right)|_{\varphi=0}=0$. The result is plotted in Fig. (\ref{FIG. 4}).

\begin{figure}
     \centering
     \begin{subfigure}[b]{0.4\textwidth}
         \centering
         \includegraphics[width=\textwidth]{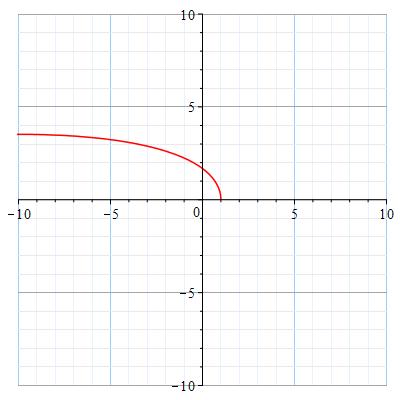}
         \caption{numeric plot of $r$ versus $\varphi$ for positive charge,$Q=2$ and small $a$}
         \label{Fig. 5a}
     \end{subfigure}
     \hfill
     \begin{subfigure}[b]{0.4\textwidth}
         \centering
         \includegraphics[width=\textwidth]{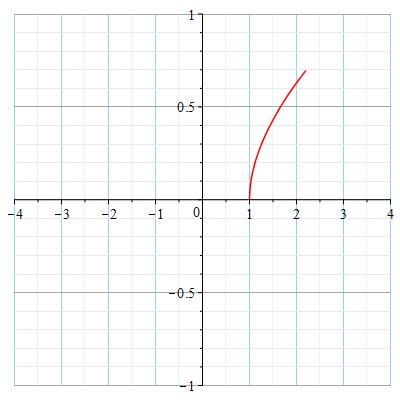}
         \caption{numeric plot of $r$ versus $\varphi$ for negative charge,$Q=-2$ and small $a$}
         \label{Fig. 5b}
     \end{subfigure}
        \caption{Numeric plot of $r$ versus $\varphi$ for both positive and negative charge. For, $l=1$, $Q= \pm2$, and $a=0.3$, the initial point $r_{0}=1$ so that for positive charge $E=1.893082259$ and for the negative charge $E=3.272268399$. Unlike Fig. (\ref{FIG. 4}), it is seen that for small rotation parameter $\left(a=0.3\right)$ the right-handed chiral behavior is no more valid.}
        \label{FIG. 5}
\end{figure}

For rotation parameter $a$, if it is considered so small between $0<a<1$ then the direction of rotation for the charged particle as is shown in Fig. (\ref{FIG. 5}) is different. Namely, for small rotation parameters ($\pm$) charges tend to incline oppositely.
\section{Existence of CTCs in RBR spacetime}
We consider the line element (\ref{1}) and apply the original technique used first by G\"{o}del in the perfect fluid model spacetime in the presence of a cosmological constant \cite{9}. Take the $r=r_{0}=$constant and $\theta=\theta_{0}=$constant in the metric with $t=-\alpha\phi$, where $\alpha$ is a positive small constant parameter. $\phi$ is a periodic angle with period $2\pi$. Identification of $\phi=0$ and $\phi=2\pi$ will reflect on the time variable to carry $t$ backward in time. Whether this is possible or not will be checked from the metric (\ref{1}). Applying
\begin{align}
     r & =r_{0}=\textit{const.} \nonumber\\
    \theta & =\theta_{0}=\textit{const.}\nonumber \\
    t & =-\alpha\phi \label{32}
\end{align}
will lead us to 
\begin{equation}
    ds^{2}=\left[\alpha^{2}\left(\frac{F^{2}}{\sin^{2}\theta_{0}}-q^{2}\right)-r_{0}^{2}-2\alpha q r_{0}\right]d\phi^{2}\label{33}
\end{equation}
which must be positive for timelike curves. This amounts to the inequality
\begin{equation}
    \alpha^{2}\left(\frac{F^{2}}{\sin^{2}\theta_{0}}-q^{2}\right)>r_{0}^{2}+2\alpha q r_{0}\label{34}
\end{equation}
which must hold true. First, the expression 
\begin{equation}
    \frac{F^{2}}{\sin^{2}\theta_{0}}-q^{2}>0\label{35}
\end{equation}
must be satisfied. We test this for the two particular cases: for $\theta_{0}=\frac{\pi}{2}$ and $\theta_{0}=\frac{\pi}{4}$, as follows\\
i)  $\theta_{0}=\frac{\pi}{2}$ case.
\begin{equation}
   \frac{F^{2}}{\sin^{2}\theta_{0}}-q^{2}=\left(1+a^{2}\right)^{2}-4a^{2}\left(1+a^{2}\right)>0\nonumber
\end{equation}
which gives that $a^{2}<\frac{1}{3}$. Choosing $a^{2}=\frac{1}{4}$ and $\alpha=\frac{1}{\sqrt{5}}$ provides us that the inequality (\ref{34}) holds true for the interval $0<r_{0}<\frac{1}{4}\left(\sqrt{2}-1\right)$.\\
ii) $\theta_{0}=\frac{\pi}{4}$ case\\
The inequality (\ref{35}) holds true for all values of $a$ parameter in this case. Then, the choice $a=1$ and $\alpha=\frac{1}{2\sqrt{2}}$ leads us to the interval $0<r_{0}<\sqrt{\frac{17}{8}}-1$, which makes the inequality (\ref{34}) satisfied. For any other choice of the parameter $a$, and a choice for $\alpha$ also will give us an appropriate interval in which the critical inequality (\ref{34}) satisfies.\par
These two examples for  $\theta_{0}=\frac{\pi}{2}$ and  $\theta_{0}=\frac{\pi}{4}$ can be extended to any angular variable in which a possible domain for the $r_{0}$ variable will be available. It is observed that the $r_{0}$ range emerges rather confined. The same argument can be extended to apply to the different metric representations. Namely, for any $\theta_{0}$, the parameters $\alpha$, $a$ can be tuned to provide an interval for $r=r_{0}$ to provide the required inequality (\ref{34}). In short, the existence of the rotation parameter in the RBR spacetime makes it possible to go backward in time in a periodic manner to make CTCs. What is interesting, as our simple analysis shows, CTCs exist in the NHG of any charged, spinning BH.

\section{Conclusion and discussion}
There has been great interest in recent times about the shadow of Sagittarius *A at the center of our galaxy. In particular, the electromagnetic (em) patterns around the event horizon (EH) of the supermassive BH, polarization of the fields, and jet ejection are of utmost importance. Our spacetime, the RBR was proved in the past and repeated briefly herein to represent the near horizon geometry (NHG) of the Kerr-Newman (KN) BH. The distribution of $\Vec{E}$ and $\Vec{B}$ fields in RBR therefore corresponds to the fields in the vicinity of KN BH. Spinning of the BH polarizes the em fields. Geodesics of charged particles are solved, representing right-handed curves for finite rotations irrespective of the sign of charges. For weak rotations, this effect is not observed. In the case of fast rotations of the BH, the $\Vec{B}$ field happens to be much weaker. It remains to be seen whether our exact results have correspondence in the spinning- $\Vec{B}$ field connection in astrophysics. We argue that the right-handedness of the geodesic flow can be considered to define a chirality at a cosmic level. Finally, since closed timelike curves (CTCs) arise in any spinning metric the RBR spacetime also can not be an exception. Such CTCs, however, arise as highly localized micro-circular paths. Our analysis suggests that CTCs are a natural phenomenon not only in em cosmos, but also in the NHG of all spinning BHs. Further, the occurrence of CTCs is based on a metric powered by a physical energy-momentum, not by exotic matter as in the case of a wormhole.

\end{document}